# Identifying Speakers Using Their Emotion Cues

Ismail Shahin

Electrical and Computer Engineering Department

University of Sharjah

P. O. Box  27272

Sharjah, United Arab Emirates

Tel: (971) 6 5050967

Fax: (971) 6 5050877

E-mail: ismail@sharjah.ac.ae

**Abstract**

This paper addresses the formulation of a new speaker identification approach which employs knowledge of emotional content of speaker information. Our proposed approach in this work is based on a two-stage recognizer that combines and integrates both emotion recognizer and speaker recognizer into one recognizer. The proposed approach employs both Hidden Markov Models (HMMs) and Suprasegmental Hidden Markov Models (SPHMMs) as classifiers. In the experiments, six emotions are considered including neutral, angry, sad, happy, disgust and fear. Our results show that average speaker identification performance based on the proposed two-stage recognizer is 79.92% with a significant improvement over a one-stage recognizer with an identification performance of 71.58%. The results obtained based on the proposed approach are close to those achieved in subjective evaluation by human listeners.

## 1. Introduction

Speaker recognition focuses on extracting, characterizing and recognizing the information in speech signals conveying speaker identity. There are two kinds of speaker recognition: speaker identification and speaker verification (authentication). Speaker identification decides who is speaking from a set of known speakers, whereas speaker authentication determines whether a speaker belongs to a particular known voice or to some other unknown voice. Speaker recognition is divided into two sets: "open set" and "closed set". In the "open set", a reference model for the unknown speaker may not exist. However, in the "closed set", a reference model for the unknown speaker should be available to the system. Speaker recognition typically functions in one of two styles: text-dependent (fixed-text) style or text-independent (free-text) style. Text-dependent requires a user to regenerate utterances containing the same text. In the text-independent, there is no prior knowledge of the text to be spoken.

It is well known that speaker recognition performance is almost ideal in neutral talking environments [1], [2], [3], [4]; on the other hand, the performance is sharply degraded in emotional talking environments [5], [6], [7], [8], [9]. In this work, we address the issue of enhancing the degraded speaker identification performance in emotional talking environments by proposing, implementing and testing a new approach. This approach is based on identifying the unknown speaker using his/her emotion cues.

## 2. Motivation

In literature, there are some studies that focus on the field of speaker recognition in emotional talking environments. Bao *et al*. focused in one of their studies on emotion attribute projection for speaker recognition on emotional speech [5]. Wu *et al.* investigated the rules based feature modification for robust speaker recognition with emotional speech [6]. Li *et al.* proposed an approach of speech emotion-state conversion to enhance speaker recognition performance in emotional talking environments [7]. Shahin focused in two of his earlier studies on using emotions to identify speakers (emotion-dependent speaker identification) [8] and on speaker identification in emotional talking environments [9]. In the first study, he achieved an average speaker identification performance of 78.8% (in a closed set with forty speakers and six emotions) [8]. In the second study, he obtained an average speaker identification performance of 61.4%, 66.4% and 69.1% based on, respectively, hidden Markov models, second-order circular hidden Markov models and suprasegmental hidden Markov models using forty speakers and five emotions [9].

The contribution of this work is positioned on proposing, implementing and evaluating a new approach based on a two-stage recognizer in emotional talking environments. The two-stage recognizer combines and integrates both emotion recognizer and speaker recognizer into one recognizer in a closed set of text-independent using both HMMs and SPHMMs as classifiers. The aim of the new proposed approach is to alleviate the deteriorated speaker identification performance in such talking environments. The emotional talking environments in this work consist of six emotions including the neutral state. These emotions are neutral, angry, sad, happy, disgust and fear. The applications of speaker identification in emotional talking environments appear in criminal investigations to identify the suspected persons who produced voice in such talking environments and in Text-To-Speech (TTS) communication-aid that can help expressing the emotions of the speaker. Traditional techniques that convert text to speech result in a dry message. However, adding emotion (e.g. happy emotion or angry emotion) to the read text would result in a more realistic and live message.

The structure of the paper is as follows. The overviews of SPHMMs are given in the next section. Section 4 describes the speech database used to assess the proposed approach. Section 5 is committed to discussing the proposed approach and the experiments. Section 6 discusses the results achieved in this work. Concluding remarks are given in Section 7.

## 3. Overviews of Suprasegmental Hidden Markov Models

Many classifiers have been proposed in the areas of speech recognition and speaker recognition including Gaussian Mixture Models (GMMs) [10], Artificial Neural Networks (ANNs) [1], HMMs [11], [12], [13] and Support Vector Machines (SVMs) [14].

Suprasegmental hidden Markov models have been developed, implemented and tested for speaker identification in shouted talking condition [15] and in emotional talking environments [9]. SPHMMs have proven to be superior models over HMMs for speaker recognition systems in each of the emotional and shouted talking environments [9], [15].

Several states of HMMs can be encapsulated into a new state called suprasegmental state. Suprasegmental state can look at the observation sequence through a larger window. Such a state permits observations at proper rates for the situation of modeling. As an example, prosodic information can not be observed at a rate that is used for acoustic modeling. The most important acoustic parameters that express prosody are fundamental frequency, intensity and duration of speech signals [16]. The prosodic features of a unit of speech are labeled suprasegmental features because they have impact on all the segments of a speech signal unit. Therefore, prosodic events at the stages of phone, syllable, word and utterance are represented using suprasegmental states; on the other hand, acoustic events are represented using conventional hidden Markov states.

Prosodic information can be combined with acoustic information within HMMs [17]. The following formula shows how to perform this combination,

$$log\ P(\lambda^v, \Psi^v | O) = (1-\alpha).\ log\ P(\lambda^v | O) + \alpha.\ log\ P(\Psi^v | O) \quad (1)$$

where $\alpha$ is a weighting factor. When:

$$\begin{cases} 0.5 > \alpha > 0 & \text{biased towards acoustic model} \\ 1 > \alpha > 0.5 & \text{biased towards prosodic model} \\ \alpha = 0 & \text{biased completely towards acoustic model and} \\ & \text{no effect of prosodic model} \\ \alpha = 0.5 & \text{no biasing towards any model} \\ \alpha = 1 & \text{biased completely towards prosodic model and} \\ & \text{no impact of acoustic model} \end{cases} \quad (2)$$

$\lambda^v$: is the acoustic model for the $v$th emotion.

$\Psi^v$: is the suprasegmental model for the $v$th emotion.

$O$: is the observation vector or sequence of an utterance.

$P(\lambda^v | O)$: is the probability of the $v$th HMM emotion model given the observation vector $O$.

$P(\Psi^v | O)$: is the probability of the $v$th SPHMM emotion model given the observation vector $O$. *Refs.* [9] and [15] have more information about suprasegmental hidden Markov models.

# 4. Speech Database

A total of twenty five male speakers and twenty five female speakers were asked to generate the speech database used to evaluate the new proposed approach. All the speakers were healthy adult native speakers of American English. The speakers were asked to portray eight sentences nine times each under each of the neutral, angry, sad, happy, disgust and fear emotions. The first four sentences were used in the training stage, while the last four sentences were used in the test stage (text-independent experiment). The total number of utterances was 21600 (50 speakers times 8 sentences times 9 utterances/sentence times 6 emotions). The eight sentences were unbiased towards any emotion (no correlation between any sentence and any emotion). The eight sentences are:

1) *He works five days a week.*
2) *The sun is shining.*
3) *The weather is fair.*
4) *The students study hard.*
5) *Assistant professors are looking for promotion.*
6) *University of Sharjah.*
7) *Electrical and Computer Engineering Department.*
8) *He has two sons and two daughters.*

A speech acquisition board with a 16-bit linear coding A/D converter and a sampling rate of 16 kHz was used to capture the speech database in an uncontaminated environment. The database was a 16-bit per sample linear data. The speech signals were applied every 5 ms to a 30 ms Hamming window.

In this work, Mel-Frequency Cepstral Coefficients (MFCCs) have been adopted as the features to represent the phonetic content of speech signals. MFCCs have been employed in the areas of speech recognition in stressful talking environments and speaker recognition in stressful talking environments because such coefficients outperform other features in the two areas and because they offer a high-level approximation of human auditory perception [5], [18], [19]. In each of HMMs and SPHMMs, a 16-dimension feature analysis of MFCC was used to construct the observation vectors. The number of conventional states, $N$, was nine and the number of suprasegmental states was three (each three conventional states were combined into one suprasegmental state) in SPHMMs and a continuous mixture observation density was selected for each model. The number of mixture components, $M$, was ten per state.

In the last four decades, the majority of research carried out in the fields of speech recognition and speaker recognition on HMMs have been done using left-to-right hidden Markov models (LTRHMMs) because phonemes follow strictly the left to right sequence [11], [20], [21]. In this work, left-to-right suprasegmental hidden Markov models (LTRSPHHMs) have been derived from LTRHMMs. Figure 1 shows an example of a basic structure of LTRSPHMMs that has been derived from LTRHMMs. In this figure, $q_1, q_2,..., q_6$ are hidden Markov states. $p_1$ is a suprasegmental state that consists of $q_1$, $q_2$ and $q_3$. $p_2$ is a suprasegmental state that is made up of $q_4$, $q_5$ and $q_6$. $p_3$ is a suprasegmental state that is composed of $p_1$ and $p_2$. $a_{ij}$ is the transition probability between the $i$th hidden Markov state and the $j$th hidden Markov state, while $b_{ij}$ is the transition probability between the $i$th suprasegmental state and the $j$th suprasegmental state. The transition matrix, $A$, of this structure using the two suprasegmental states $p_1$ and $p_2$ can be expressed in terms of the positive coefficients $b_{ij}$ as,

$$A = \begin{bmatrix} b_{11} & b_{12} \\ 0 & b_{22} \end{bmatrix}$$

## 5. Emotion-Dependent Speaker Identification Approach and the Experiments

Given *n* speakers talking in *m* emotions, the proposed architecture is composed of two cascaded stages as illustrated in Figure 2. This figure shows that emotion-dependent speaker identification recognizer is nothing but a two-stage recognizer that integrates and combines both emotion recognizer and speaker recognizer into one system. The two stages are:

**Stage *a*: Emotion Recognizer**

First, the emotion of the unknown speaker is identified (emotion identification problem). In this stage, *m* probabilities are computed based on SPHMMs and the maximum probability is chosen as the identified emotion as given in the following formula,

$$E^* = \arg\max_{m \geq e \geq 1} \left\{ P\left(O \mid \lambda^e, \Psi^e\right) \right\} \quad (3)$$

where,

$E^*$: is the index of the identified emotion.

*O*: is the observation sequence of the unknown emotion that belongs to the unknown speaker.

$P\left(O \mid \lambda^e, \Psi^e\right)$: is the probability of the observation sequence *O* of the unknown emotion that belongs to the unknown speaker given the *e*th SPHMM emotion model.

In this stage, the *e*th SPHMM emotion model has been obtained in the training session for every emotion using the fifty speakers uttering all the first four sentences of the database (text-independent) with a repetition of nine utterances/sentence. The total number of utterances used to derive each SPHMM emotion model in this session is 1800 (50 speakers times 4 sentences times 9 utterances/sentence). The training session of SPHMMs is very similar to the training session of the conventional HMMs. In the training session of SPHMMs, suprasegmental models are trained on top of acoustic models of HMMs. A block diagram of this stage is shown in Figure 3. Derivation of the *e*th SPHMM emotion model in this training session is illustrated in Figure 4.

**Stage *b*: Speaker Recognizer**

Given that the emotion of the unknown speaker was identified, the next stage is to identify the unknown speaker (emotion-specific speaker identification problem). In this stage, *n* probabilities per emotion are computed based on HMMs and the maximum probability is chosen as the identified speaker for the recognized emotion as given in the following formula,

$$S^* = \arg\max_{n \geq s \geq 1} \left\{ P\left(s \mid \lambda^S, E^*\right) \right\} \qquad (4)$$

where,

$S^*$: is the index of the identified speaker.

$P\left(s \mid \lambda^S, E^*\right)$: is the probability of the observation sequence *s* that belongs to the unknown speaker given the *s*th HMM speaker model and the identified emotion.

The *s*th HMM speaker model has been derived using nine utterances per sentence (the first four sentences of the database). The total number of utterances used to build each emotion-dependent HMM speaker model is 36 (4 sentences times 9 utterances/sentence). Derivation of the $s$th HMM speaker model for every emotion in this training session is shown in Figure 5.

In the test or identification session (completely separate from the training session), each one of the fifty speakers used nine utterances per sentence (the last four sentences of the database) under each emotion including the neutral state. The total number of utterances used in this session is 10800 (50 speakers times 4 sentences times 9 utterances/sentence times 6 emotions). This stage can be shown in the block diagram of Figure 6.

## 6. Results and Discussion

In this work, emotion cues have been used to identify the unknown speaker in emotional talking environments in order to improve the deteriorated speaker identification performance. Table 1 illustrates a confusion matrix that represents the percentage of confusion of the unknown emotion with the other emotions of stage $a$ of the proposed approach when the weighting factor ($\alpha$) is equal to 0.5. This specific value of the weighting factor has been selected to avoid biasing towards either acoustic or prosodic model. The average emotion identification performance based on this table is 83.83%. This table indicates the following:

a) The most easily recognizable emotion is neutral (94%). Therefore, the expected highest speaker identification performance will occur when speakers speak in a neutral state.

b) The least easily recognizable emotion is angry (78%). Consequently, the predicted least speaker identification performance will happen when speakers speak in an angry emotion.

c) Column 3 (angry column), for example, shows that 4% of the utterances that were portrayed in an angry emotion were evaluated as generated in a neutral state, 5% of the utterances that were uttered in an angry emotion were identified as produced in a sad emotion. This column shows that angry emotion has the highest confusion percentage with disgust emotion (10%). Therefore, angry emotion is highly confusable with disgust emotion. This column also shows that angry emotion has no confusion with happy emotion (0%).

Table 2 shows speaker identification performance based on emotion-dependent speaker identification approach (two-stage recognizer) using SPHMMs when $\alpha = 0.5$, while Table 3 gives speaker identification performance based on emotion-independent speaker identification approach (one-stage recognizer) using the same models. The average speaker identification performance based on the two-stage recognizer using SPHMMs is 79.92%, while the average speaker identification performance based on the one-stage recognizer using the same models is 71.58%.

A statistical significance test has been carried out to investigate whether speaker identification performance differences (speaker identification performance based on the two-stage recognizer and that based on the one-stage recognizer in emotional talking environments) are real or simply due to statistical fluctuations.

The statistical significance test has been performed based on the Student $t$ Distribution test as given by the following formula,

$$t_{\text{two-stage, one-stage}} = \frac{\overline{x}_{\text{two-stage}} - \overline{x}_{\text{one-stage}}}{SD_{\text{pooled}}} \quad (5)$$

where,

$\overline{x}_{\text{one-stage}}$: is the mean of the first sample (one-stage recognizer) of size $n$.

$\overline{x}_{\text{two-stage}}$: is the mean of the second sample (two-stage recognizer) of the same size.

$SD_{\text{pooled}}$: is the pooled standard deviation of the two samples (recognizers) given as,

$$SD_{\text{pooled}} = \sqrt{\frac{SD^2_{\text{one-stage}} + SD^2_{\text{two-stage}}}{n}} \quad (6)$$

where,

$SD_{\text{one-stage}}$: is the standard deviation of the first sample (one-stage recognizer) of size $n$.

$SD_{\text{two-stage}}$: is the standard deviation of the second sample (two-stage recognizer) of the same size.

Based on Table 2 and Table 3, $\overline{x}_{\text{one-stage}} = 71.58$, $SD_{\text{one-stage}} = 7.57$, $\overline{x}_{\text{two-stage}} = 79.92$, $SD_{\text{two-stage}} = 6.03$. Based on these values, the calculated $t$ value is $t_{\text{two-stage, one-stage}} = 6.093$. This calculated $t$ value is much greater than the tabulated critical value at *0.05* significant level $t_{0.05} = 1.645$. Therefore, it is evident from Table 2 and Table 3 that the two-stage recognizer is superior to the one-stage recognizer for speaker identification in emotional talking environments.

The achieved speaker identification performance based on the proposed approach has been compared with the results obtained in some previous studies. The proposed approach in the current work yields better speaker identification performance than that reported in some prevoius studies:

1) Emotion-independent speaker identification performance attained by Shahin. Shahin obtained in one of his studies an average speaker identification performance of 69.1% in emotional talking environments based on HMMs [9]. Hence, it is evident that inserting an emotion identification stage into emotion-independent speaker identification system significantly enhances speaker identification performance in such talking environments.

2) Three experiments designed by Li *et al.* to evaluate speaker identification performance based on their three proposed approaches. Their experiments were conducted using the Emotional Prosody Speech and Transcript corpus from the Linguistic Data Consortium (LDC) [22]. This database consists of 8 professional speakers (three actors and five actresses) talking in 14 emotional states in addition to the neutral state. These proposed approaches are [7]:

   i. Baseline approach. This approach is based on using MFCCs as the features and Gaussian Mixture Models (GMMs) as a classifier. Based on this approach, they reported 62.81% as an average speaker identification performance.

   ii. Unmodified Linear Predictive Coding (LPC) approach. This approach is based on LPC analysis and synthesis. The attained average speaker identification performance based on this approach was 62.34%.

iii. Emotion-state conversion approach. This approach was proposed to enhance speaker identification performance in emotional talking conditions. They achieved 70.22% as an average speaker identification performance based on their proposed approach.

Speaker identification performance using his/her emotions based on the proposed approach is limited as shown in Table 2. This table is the resultant of both stage *a* and stage *b*. The reasons of limitations are:

i. The unknown emotion that belongs to the unknown speaker in stage *a* is not perfectly identified. The emotion identification performance of this stage as calculated from Table 1 is 83.83%.

ii. The unknown speaker in stage *b* is not 100% correctly identified.

Four more experiments have been separately performed to evaluate the results achieved based on the current proposed approach. The four experiments are:

1) Experiment 1: HMMs have been employed in both stage *a* and stage *b* of the two-stage recognizer. Table 4 yields speaker identification performance based on using HMMs in both stages of the proposed recognizer. The average speaker identification performance based on Table 4 is 75.92% (with a standard deviation of 6.44). To make a comparison between a two-stage speaker identification performance based on SPHMMs and that based on HMMs, the $t_{\text{two-stage (SPHMMs), two-stage (HMMs)}}$ has been calculated. The calculated *t* value is $t_{\text{two-stage (SPHMMs), two-stage (HMMs)}}$ = 3.206. This calculated *t* value is greater than the tabulated critical value $t_{0.05}$ = 1.645. Therefore, the conclusion that can be drawn in this experiment shows that suprasegmental hidden Markov models outperform hidden Markov models for speaker identification in emotional talking environments based on the two-stage recognizer.

2) Experiment 2: Emotional Prosody Speech and Transcripts database has been used instead of the collected database. Emotional Prosody Speech and Transcripts database was generated by the Linguistic Data Consortium (LDC) to evaluate experiments carried out in emotional talking environments. This database consists of recordings captured from a limited number of speakers (three professional actors and five professional actresses). The eight speakers read a series of semantically neutral utterances that are made of dates and numbers spoken in fifteen different emotions including the neutral state [22]. Only six basic emotions, namely neutral, hot anger, sadness, happiness, disgust and panic have been used in this experiment. Table 5 demonstrates a confusion matrix based on SPHMMs using this database. This table yields average emotion identification performance of 82.67%. It is apparent that the average emotion identification performance calculated based on Table 1 is close to that calculated based on Table 5.

Speaker identification performance based on the proposed two-stage approach using SPHMMs when α = 0.5 and using Emotional Prosody database is given in Table 6. Based on this table, the average speaker identification performance is 78.92%. It is evident from Table 2 and Table 6 that the two average speaker identification performances are close to each other.

3) Experiment 3: The proposed two-stage recognizer has been assessed for different values of the weighting factor ($\alpha$). Figure 7 shows speaker identification performance based on the proposed approach for different values of $\alpha$ (0.0, 0.1, 0.2, …, 0.9, 1.0). This figure demonstrates that increasing the value of the weighting factor has a significant effect on enhancing speaker identification performance in emotional talking environments (excluding neutral state) based on the proposed approach. Based on the two-stage recognizer, it is apparent that suprasegmental hidden Markov models have more impact on speaker identification performance than acoustic hidden Markov models in such talking environments.

4) Experiment 4: An informal subjective assessment of the proposed two-stage approach was conducted using the collected speech database with ten nonprofessional listeners (human judges). A total of 1200 utterances (fifty speakers times six emotions times four sentences only) were used in this assessment. During the evaluation, the listeners were asked to answer two questions for every test utterance. The two questions were: identify the unknown emotion and identify the unknown speaker. The results of the evaluation were satisfactory and encouraging. The average emotion identification performance was 85.02% and the average speaker identification performance was 81.01%. These averages are close to the achieved averages based on the proposed two-stage approach.

## 7. Concluding Remarks

In the present work, we proposed, applied and tested a new approach based on using emotion cues to enhance speaker identification performance in emotional talking environments. This work showed that the significant improvement of speaker identification performance using the proposed two-stage recognizer over the one-stage recognizer demonstrated the promising results of the proposed approach. Therefore, emotion cues significantly contribute in alleviating the declined speaker identification performance in emotional talking environments. This work also demonstrated that suprasegmental hidden Markov models are superior to hidden Markov models for speaker identification in such talking environments. Speaker identification performance based on emotion cues is limited. The performance of the overall proposed approach is the resultant of the performances of:

a) Emotion identification stage. The emotion of the unknown speaker in this stage is not 100% correctly identified.

b) Speaker identification stage. The unknown speaker in this stage is not perfectly identified.

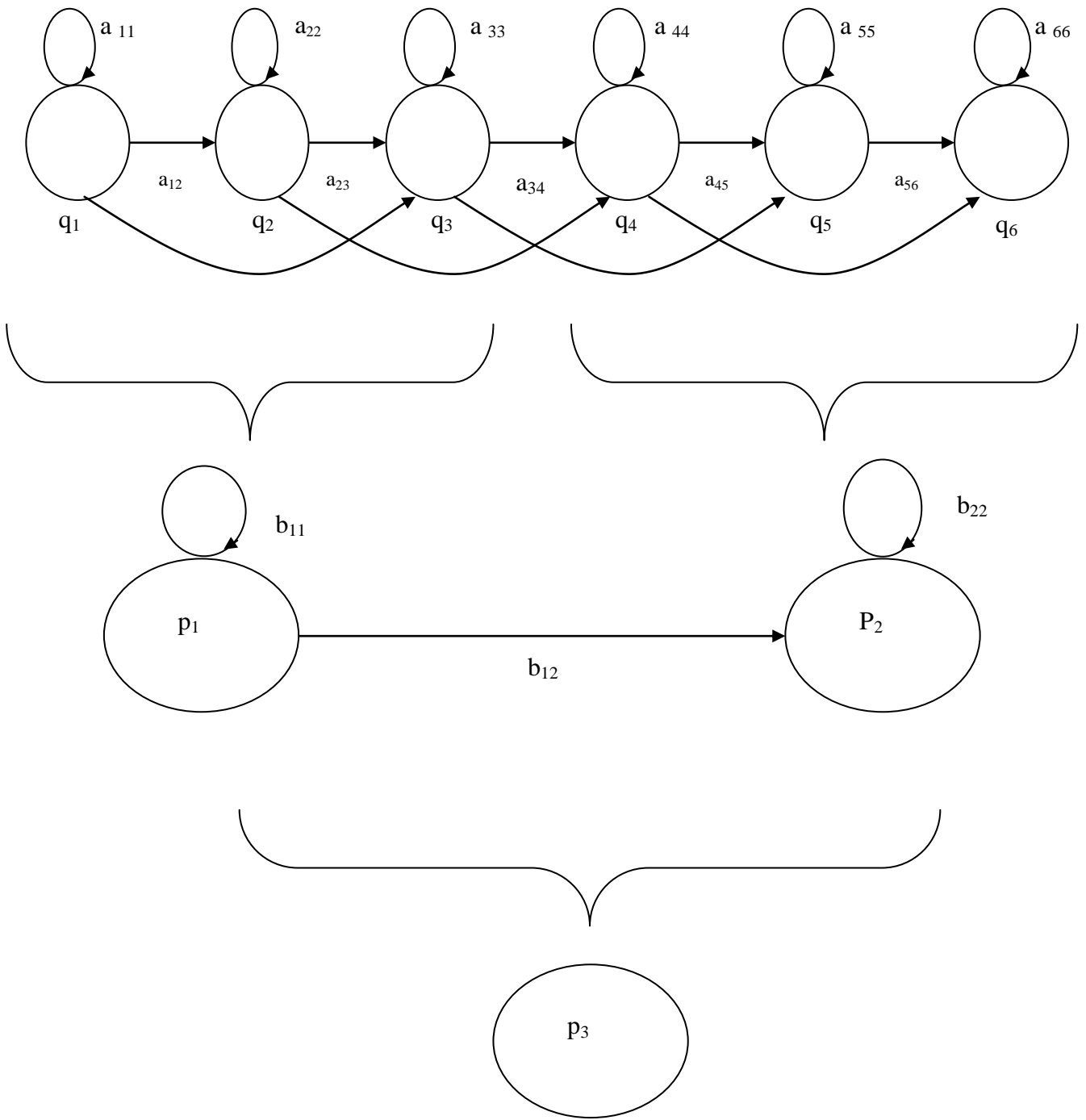

**Figure 1.** Basic structure of LTRSPHMMs derived from LTRHMMs

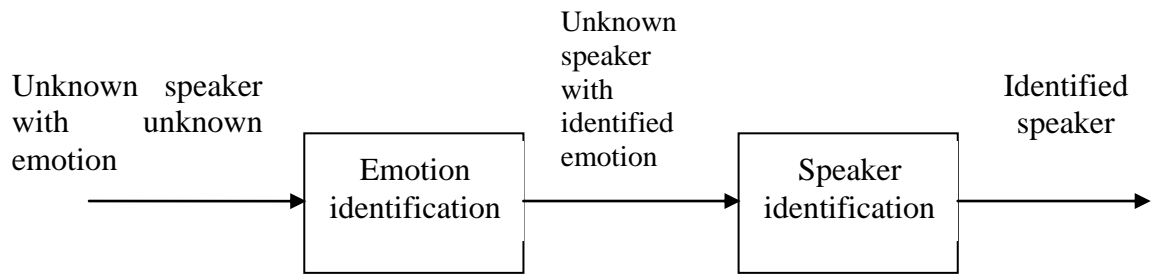

**Figure 2.** Block diagram of the overall proposed two-stage recognizer

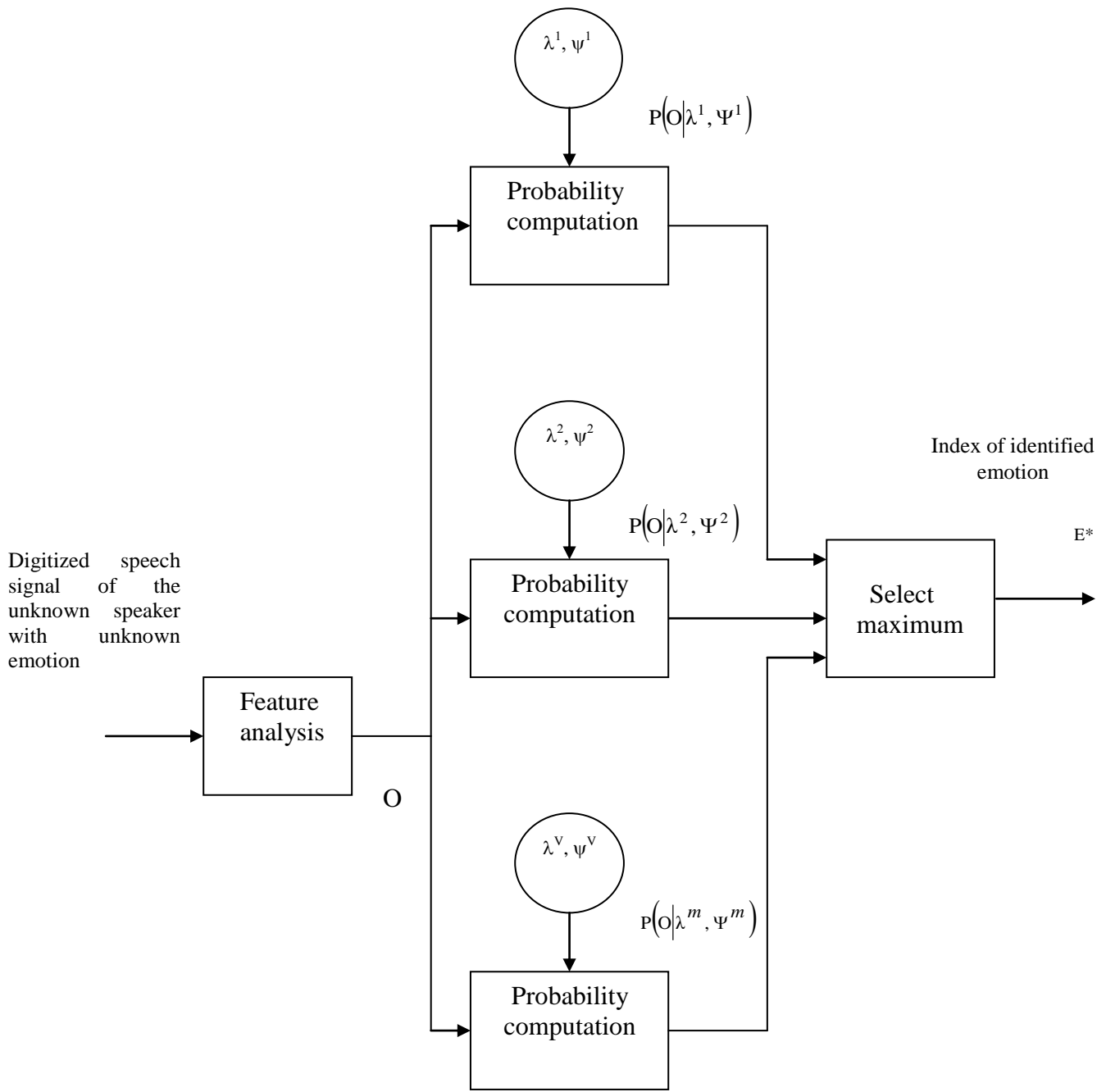

**Figure 3.** Block diagram of stage *a* of the proposed approach based on SPHMMs

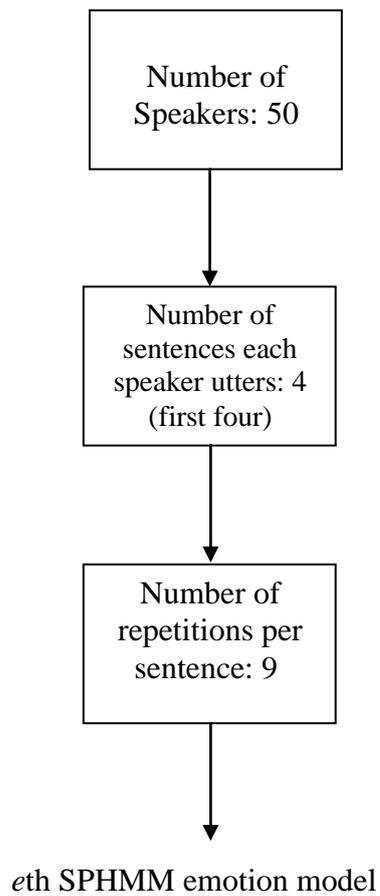

**Figure 4.** Derivation of the *e*th SPHMM emotion model in the training session of stage *a* of the proposed approach

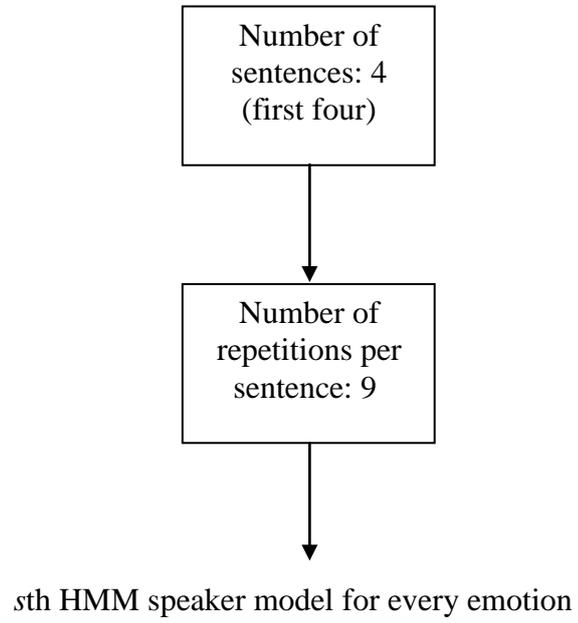

*s*th HMM speaker model for every emotion

**Figure 5.** Derivation of the *s*th HMM speaker model for every emotion in the training session of stage *b* of the proposed approach

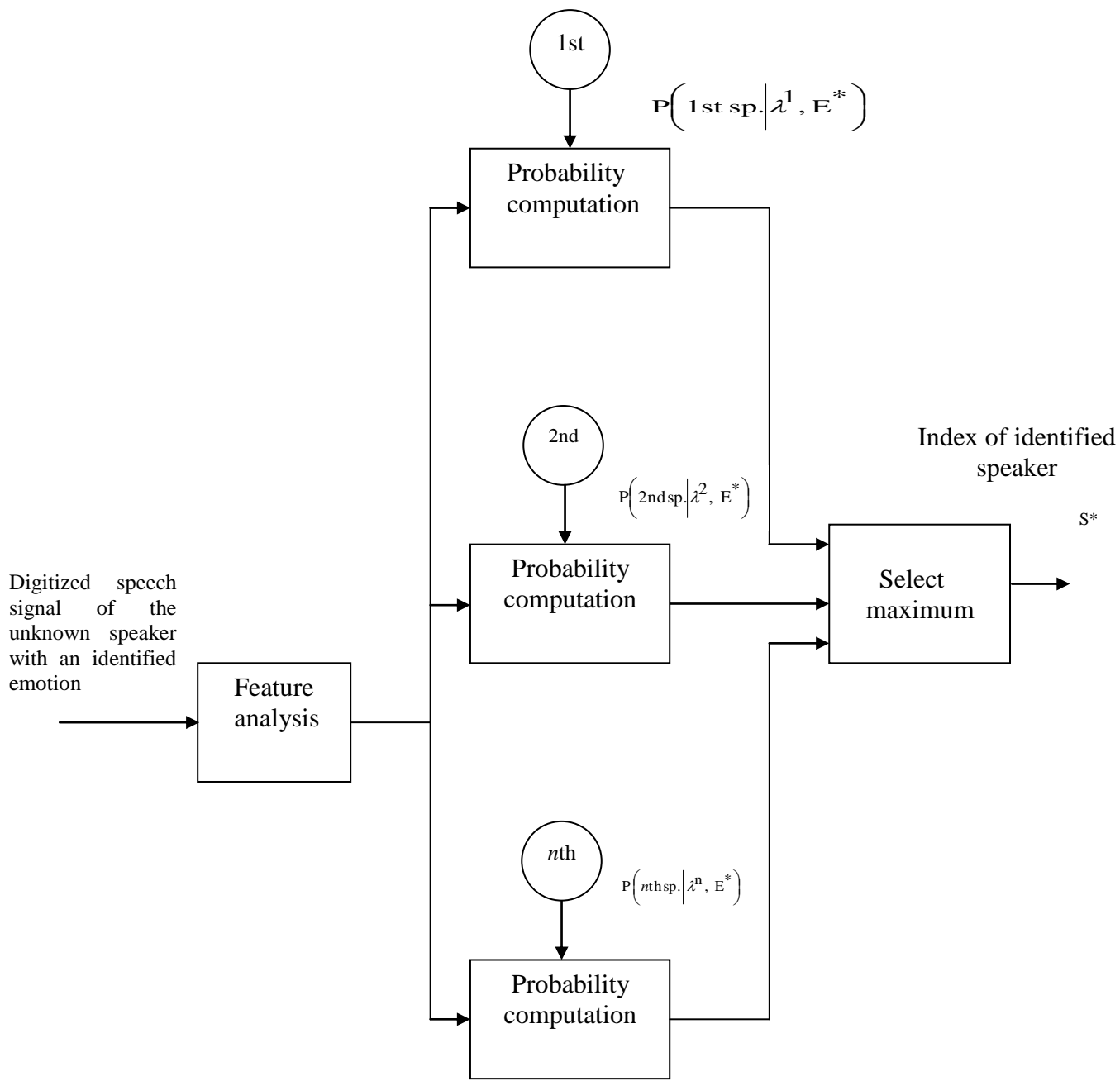

**Figure 6.** Block diagram of stage *b* of the proposed approach based on HMMs

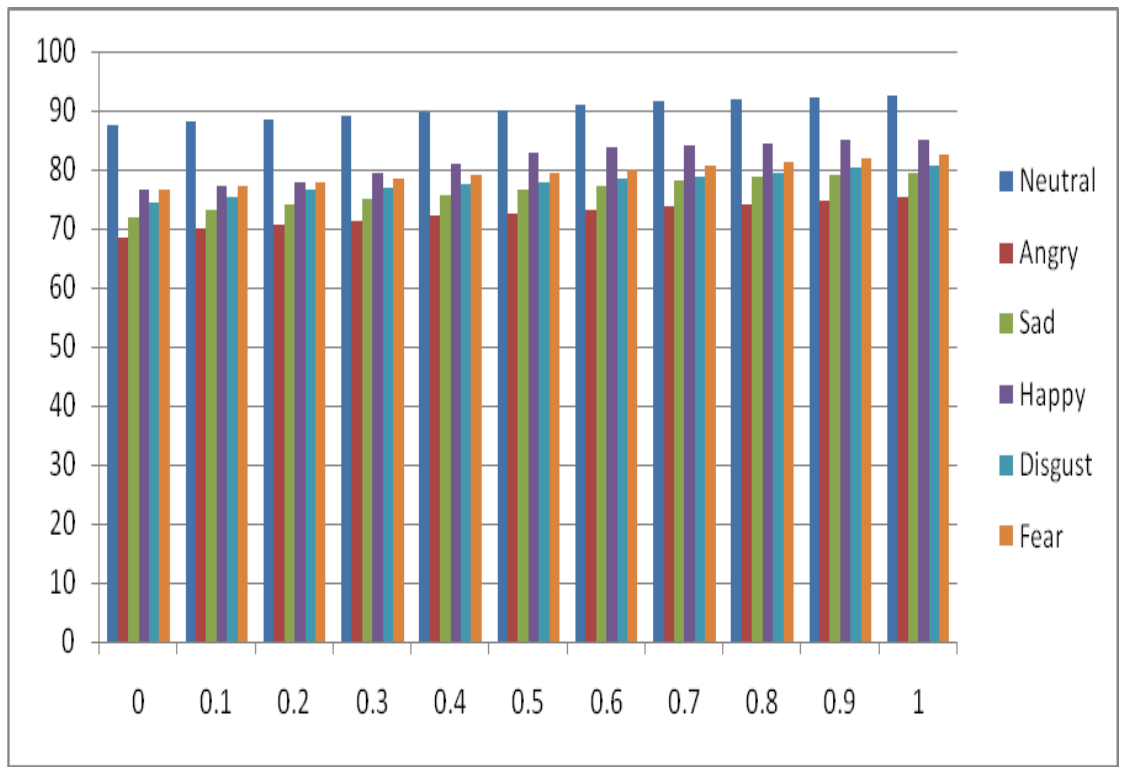

**Figure 7.** Speaker identification performance (%) versus the weighting factor ($\alpha$) based on the proposed approach

Table 1

Confusion matrix based on stage *a* of the proposed approach using SPHMMs when α = 0.5

| | Percentage of confusion of a test emotion with the other emotions | | | | | |
|---|---|---|---|---|---|---|
| Model | Neutral | Angry | Sad | Happy | Disgust | Fear |
| Neutral | **94%** | 4% | 2% | 4% | 2% | 2% |
| Angry | 0% | **78%** | 6% | 2% | 10% | 3% |
| Sad | 4% | 5% | **80%** | 2% | 3% | 7% |
| Happy | 1% | 0% | 2% | **88%** | 1% | 2% |
| Disgust | 0% | 10% | 2% | 1% | **80%** | 3% |
| Fear | 1% | 3% | 8% | 3% | 4% | **83%** |

Table 2

Speaker identification performance based on the two-stage recognizer using

SPHMMs when $\alpha = 0.5$

| Emotion | Males (%) | Females (%) | Average (%) |
|---------|-----------|-------------|-------------|
| Neutral | 89 | 91 | 90 |
| Angry | 72 | 73 | 72.5 |
| Sad | 76 | 77 | 76.5 |
| Happy | 82 | 84 | 83 |
| Disgust | 78 | 78 | 78 |
| Fear | 80 | 79 | 79.5 |

Table 3

Speaker identification performance based on the one-stage recognizer using SPHMMs

| Emotion | Males (%) | Females (%) | Average (%) |
|---|---|---|---|
| Neutral | 84 | 86 | 85 |
| Angry | 62 | 62 | 62 |
| Sad | 67 | 69 | 68 |
| Happy | 73 | 72 | 72.5 |
| Disgust | 71 | 70 | 70.5 |
| Fear | 71 | 72 | 71.5 |

Table 4

Speaker identification performance based on the two-stage recognizer using HMMs in both stages

| Emotion | Males (%) | Females (%) | Average (%) |
|---|---|---|---|
| Neutral | 87 | 88 | 87.5 |
| Angry | 68 | 69 | 68.5 |
| Sad | 71 | 73 | 72 |
| Happy | 76 | 77 | 76.5 |
| Disgust | 74 | 75 | 74.5 |
| Fear | 77 | 76 | 76.5 |

Table 5

Confusion matrix based on stage *a* of the proposed approach using SPHMMs and Emotional Prosody database

| | Percentage of confusion of the unknown emotion with the other emotions | | | | | |
|---|---|---|---|---|---|---|
| Model | Neutral | Hot Anger | Sad | Happy | Disgust | Panic |
| Neutral | **96%** | 3% | 3% | 7% | 1% | 1% |
| Hot Anger | 0% | **75%** | 6% | 2% | 8% | 2% |
| Sad | 1% | 5% | **77%** | 2% | 5% | 8% |
| Happy | 1% | 3% | 2% | **84%** | 1% | 3% |
| Disgust | 0% | 8% | 4% | 2% | **82%** | 4% |
| Panic | 2% | 6% | 8% | 3% | 3% | **82%** |

Table 6

Average speaker identification performance based on the proposed two-stage approach using SPHMMs when $\alpha = 0.5$ and using Emotional Prosody database

| Emotion | Average (%) |
|---------|-------------|
| Neutral | 91.5 |
| Hot Anger | 71.5 |
| Sad | 74 |
| Happy | 82.5 |
| Disgust | 76.5 |
| Panic | 77.5 |